\documentclass[twocolumn,showpacs,preprintnumbers,amsmath,amssymb]{revtex4}
\usepackage{graphicx}
\usepackage{dcolumn}
\usepackage{bm}
\begin{document}
\preprint{APS/123-QED}
\title{THE MECHANICS OF THE SYSTEMS OF STRUCTURED PARTICLES AND IRREVERSIBILITY}
\author{V.M. Somsikov}
 \altaffiliation[] {}
 \email{nes@kaznet.kz}
\affiliation{%
Laboratory of Physics of the geoheliocosmic relation, Institute of
Ionosphere, Almaty, Kazahstan.
}%

\date{\today}
\begin{abstract}
Dynamics of systems of structured particles consisting of
potentially interacting material points is considered in the
framework of classical mechanics. Equations of interaction and
motion of structured particles have been derived. The expression for
friction force has been obtained. It has been shown that
irreversibility of dynamics of structured particles is caused by
increase of their internal energy due to the energy of motion.
Possibility of theoretical substantiation of the laws of
thermodynamics has been considered.
\end{abstract}

\maketitle

\section{\label{sec:level1}Introduction\protect}

Collective properties of natural systems are to be related to laws
of dynamics and properties of their constituent elements. Revealing
of these relations is among the main tasks of physics. There are
obstacles of principal nature towards solving this task mainly
caused by absence of connection between the laws of classical
mechanics and thermodynamics. Within this context, the
irreversibility problem formulated by Boltzmann about 150 years ago
is the key one [1-4]. The essence of this problem is in irreversible
nature of dynamics in natural systems while in Newton equation the
dynamics of elements in a system is reversible.

Boltzmann's explanations of the irreversibility are based on
extremely small probability of noticeable deviation of thermodynamic
system from equilibrium [2, 5]. Entirely, this explanation uses
statistical laws which are alien to classical mechanics. Attempts
made to find more strict substantiation of irreversibility employing
Boltzmann equation or H-theorems encountered, as a rule, Poincare's
theorem of reversibility. This theorem forbids equilibrium in
Hamiltonian systems [3]. Nevertheless, Boltzmann methods of the
kinetic theory have appeared very effective when studying
development of irreversible processes, as well as in development and
substantiation of kinetic methods for description of nonequilibrium
systems [6-12].

Aspiring to solve the irreversibility problem within classical
mechanics, Liouville has proved that only systems which by canonical
transformations of independent variables are split into
one-degree-of- freedom systems are integrable [3]. Such splitting is
only possible in the situation with no interaction between elements
of the system. At the same time Poincare proved that, as a rule,
dynamic systems are not integrable. Assuming the potentiality
condition for elements' interaction forces leads to potentiality of
forces between systems of these elements, it follows from
Hamiltonian formalism that the task could always be reduced to
integrable systems of one degree of freedom. So, we face a
contradiction: on one hand it is proved that the class of integrable
systems is very narrow, and on the other hand a priori condition
accepted for interaction forces should assure integration of any
natural systems. In spite of remaining impossibility within
Hamiltonian formalism to bind time-reversible microscopic equations
with time-irreversible macroscopic equations of thermodynamics,
symmetry still allows us to expand application of the formalism for
thermodynamics [13].

Intensive attempts to understand the essence of irreversibility were
undertaken by Planck. He has tried to prove irreversibility of
radiation (black body radiation) but he failed to overcome
Poincare's reversibility theorem. And though the irreversibility
problem remained, Planck has come to the well-known formula of a
black body equilibrium radiation, quantum of energy concept and to
creation of the quantum theory [14].

Gibbs has developed the fundament of strict statistical theory for
equilibrium systems within the frame of classical mechanics
splitting equilibrium systems into so-called ensembles. He has
thereby connected the probabilistic meaning of initial and boundary
conditions with determinism of classical mechanics. But
applicability of his theory is restricted by equilibrium systems
since the theory assumed zero energy exchange between the subsystems
[5, 15].

After Kolmogorov, Arnold and Mazur and their so-called ÊÀÌ theory
(Kolmogorov-Arnold-Ìazur), the non-integrability problem was
conventionally considered as a starting point for further
development of stochastic dynamics [3, 4, 16]. Investigations of
Hamiltonian systems and chaos due to exponential instability has led
to creation of a deterministic chaos theory. This theory has been
very fruitful. Employing the chaotic mechanics methods, the relation
between entropy and Lyapunov exponents has been revealed and there
were also studied the mechanisms of mixing and correlation splitting
[3, 17-19]. But nevertheless the problem of irreversibility has
remained unsolved due to the difficulties of the so-called "coarse
grain" of the phase space explanations.

Essential contribution to studies of irreversibility was made by I.
Prigogine and his so-called Brussels school. In particular, they
suggested and developed a method for analysis of chaotic systems
based on projection operator in Hilbert space [3, 20]. Within this
method irreversibility arose due to finite predictability horizon in
the dynamics of Hamilton systems. Like A. Poincare, I. Prigogine
also admitted possible limitation of classical mechanics [3, 21].

Trying to expand applications of statistical physics there was
developed a method of non-extensive thermodynamics used for
stationary nonequilibrium systems. The method allows to define the
distribution function for weak nonequilibrium systems and to study
relations between thermodynamic and mechanical parameters [22]. But
like any other statistical method, this method cannot be used to
explain irreversibility.

Considerable contribution to studies of nonequilibrium systems was
made by Yu. Klimontovich who has developed a statistical theory of
open systems [23]. The core of his approach verified in our research
is in consideration of continuous medium structure at all levels of
its description. Still, he has remained within statistical
description of systems.

Essential progress in solution of a problem of interrelation between
micro- and macro-processes has been achieved with a so-called
Fluctuation Theorem [24] which made it possible to describe
statistically entropy production in systems of limited number of
elements.

Considerable attention to entropy production mechanism in stationary
nonequilibrium systems was paid in [12]. It is interesting that
based on these works the phenomenological equation of motion for a
system with internal potential and external non-potential components
is currently used.

Thus, all widely known explanations of irreversibility are based on
certain hypotheses and these hypotheses go outside the constraints
of fundamental laws and principles of classical mechanics. Therefore
it is very important either to find a deterministic explanation of
irreversibility or to prove impossibility of its existence. In this
context it is interesting to consider dynamics of structured
particles when they are defined as equilibrium subsystems (ES)
consisting of potentially interacting material points. It turns out
that under certain conditions dynamics of such systems is
irreversible [25-27]. The conditions are formulated as follows:

1). The energy of an ES must be presented as a sum of internal
energy and the energy of ES motion as a whole.

2). Each material point in the system must be connected with a
certain ES independent of its motion in space.

3). During all the process the subsystems are considered to be
equilibrium.

The first condition is necessary to introduce internal energy in the
description of system dynamics as a new key parameter charactering
energy variations in ES. The second condition enables not to
redefine ES after mixing of material points. The last condition is
taken from thermodynamics. It is equivalent to the condition of weak
interactions in the ES, which do not violate ES equilibrium.
Moreover, it implies that each ES contains so many elements that it
can be described using the concept of equilibrium system.

Objectives of this paper were to analyze the dynamics of systems of
structured particles when such systems are defined as ES and to
determine the difference between their dynamics and dynamics of
Hamiltonian systems represented by potentially interacting material
points.

For that, let us derive an equitation which describes the dynamics
of two interacting equilibrium systems represented by potentially
interacting material points. We refer to this equation hereafter as
to the equation of motion for structured particles. This equation
explains the mechanism of friction in classical mechanics. It is
shown why the model of a group of ES enables to describe friction
forces. It is also shown how Lagrange, Hamilton and Liouville
equations for ES are derived from the equation of motion of
interacting ES. We consider below how such equations are different
from similar ones for systems of material points. It is shown how
classical mechanics could be linked with thermodynamics by means of
the equation of ES interaction and how the concept of entropy arises
in classical mechanics. It is shown also how based on the hypothesis
of local equilibrium, which enables to represent non-equilibrium
systems as an ensemble of ES, one can generalize the obtained
results for two interacting ES.

\section{Substantiation of the Approach}
Irreversibility is caused by dissipative processes. Therefore, if in
the framework of classical mechanics one managed to explain the
friction force and find its analytical expression; it would be
equivalent to the existence of irreversible dynamics.

In practice it is not difficult to introduce friction forces
empirically into the equation of motion, though in the frames of
classical mechanics they have not got any explanation and a
corresponding model. It is caused by the fact that classical
mechanics is based on Newton laws. Newton equation of motion gives a
relation between the changes in the body momentum and potential
forces acting on it. The friction forces are not potential forces.
They transform the motion energy of the body into the energy of
chaotic motion of elements of this body.

It is obvious that analytical description of friction forces in the
framework of classical mechanics is based on the fact that all
bodies have certain microstructure. In such a case friction may be
defined as a process of excitation of chaotic motion of a body
elements as a result of its interaction with external bodies or
field. Work of external field is spent to change internal energy of
a structured particle and to move it as determined by the systems'
CM trajectory. Part of energy converted into internal energy of the
system produces no work on motion of the system. This energy only
results in higher chaotic velocities of material points with respect
to CM. Friction forces make similar transformation of energy. So, we
would call forces which change internal energy as friction forces.
Then the description of friction is reduced to the description of
transfer of energy of relative motion of bodies into the energy of
chaotic motion of their elements as a result of action of
fundamental forces between the elements. It means that friction
forces cause changes in internal energy. It turns out that for some
models of systems one can built such description based on the law of
energy conservation. For this purpose the energy of bodies must be
presented as a sum of their energy of motion and internal energy.

We will derive the expression for the friction force using the
following method. As a model of the system we will take two
interacting structured particles moving with respect to each other.
Let us assume that each structured particle is an ES
and consists of a finite number of potentially interacting material
points. This model will enable us to combine micro-description of
the motion of material points and macro-description of motion of
ES.

To derive an equation for
transformation of the energy of motion of ES into
their internal energy it is necessary to define the energy
of each ES as a sum of internal energy and energy of
its motion. Differentiating the system energy over time and using
the condition of its conservation, we derive the equation for energy
exchange between ES and based on it determine the
equation of motion. This equation contains the friction coefficient
as a measure of transformation of the energy of system motion into
its internal energy.

\section{The equation of equilibrium systems motion}
The equation of motion for two ES can be obtained in two stages. At
the first stage, based on the condition of energy conservation we
obtain the equation of motion for the system in the field of
external forces. Then we consider a non-equilibrium system
consisting of two ES and obtain their equations of motion when the
external field for one ES is the field of forces of the other ES.
Forces acting between the ES can be obtained from their energy of
interaction.

Let us show how the equation of motion for a system of $N$ material
points with weights $m=1$ can be obtained [3-5]. As it is generally
accepted in classical mechanics, we consider forces between two
material points as potential ones. The energy of the system $E$ is
equal to the sum of kinetic energies of material points
$T_N=\sum\limits_{i=1}^{N} m{v_i}^2/2$, their potential energy in
the field of external forces, ${U_N}^{env}$, and potential energy of
their interaction
${U_N}(r_{ij})={\sum\limits_{i=1}^{N-1}}{\sum\limits_{j=i+1}^{N}}U_{ij}(r_{ij})
$, where $r_{ij}=r_i-r_j$, $r_i, v_i$ are coordinates and velocities
of the $i$-th particle. Thus, $E=E_N+U^{env}=T_N+U_N+U^{env}=const$.

By substituting variables we represent the energy of the system as a
sum of the motion energy of the CM and the
internal energy. Differentiating this energy with respect to time,
we will obtain [27]:
\begin{eqnarray}
V_NM_N\dot{V}_N+{\dot E}_N^{ins}=-V_NF^{env}-\Phi^{env}\label{eqn1}
\end{eqnarray}
Here $F^{env}=\sum\limits_{i=1}^{N}F_i^{env}(R_N,\tilde{r}_i)$,
${\dot E}_N^{ins}={\dot T}_N^{ins}(\tilde{v}_i)+{\dot
U}_N^{ins}(\tilde{r}_i)$=
$\sum\limits_{i=1}^{N}\tilde{v}_i(m\dot{\tilde{v}}_i+F(\tilde{r})_i)$,
 $\Phi^{env}=\sum\limits_{i=1}^{N}\tilde{v}_iF_i^{env}(R_N,\tilde{r}_i)$,
$r_i=R_N+\tilde{r}_i$, $M_N=mN$, $v_i=V_N+\tilde{v}_i$,
$F_i^{env}=\partial{U^{env}}/\partial{\tilde{r}_i)}$, $\tilde{r}_i$,
$\tilde{v}_i$ are the coordinates and velocity of $i$-th particle in
the CM system, $R_N,V_N$ are the coordinates and velocity of the CM
system.

The equation (1) represents the balance of the energy of the system
of material points in the field of external forces.

The first term
in the left-hand side of the equation determines the change of
kinetic energy of the system - ${\dot{T}}_N^{tr}=V_NM_N\dot{V}_N$.
The second term determines the change of internal energy of the
system, ${\dot{E}}_N^{ins}$. This energy dependent on coordinates and
velocities of material points relative to the center of mass of the
system.

The right-hand side corresponds to the work of external
forces changing the energy of the system. The first term changes
the systems motion energy. The second term determines the
work of forces changing the internal energy - ${\dot{E}}_N^{ins}$.

Let us determine the condition when the work which changes
internal energy is not equal to zero. We must take into account that
$F^{env}=F^{env}(R+\tilde{r}_i)$ where $R$ is the distance from the
source of force to the center of mass of the system. Let us assume
that $R>>\tilde{r}_i$. In this case the force $F^{env}$ can be
expanded with respect to a small parameter. Leaving in the expansion
terms of zero and first order we can write:
$F_i^{env}=F_i^{env}|_{R}+(\nabla{F_i^{env}})|_{R}\tilde{r}_i\equiv
F_{i0}^{env}+(\nabla{F_{i0}^{env}})\tilde{r}_i$. Taking into account
that $\sum\limits_{i=1}^{N}\tilde{v}_i
=\sum\limits_{i=1}^{N}\tilde{r}_i=0$ and
$\sum\limits_{i=1}^{N}F_{i0}^{env}=NF_{i0}^{env}=F_0^{env}$, we get
from (1):
\begin{eqnarray}
V_N(M_N\dot{V}_N)+
\sum\limits_{i=1}^{N}m\tilde{v}_i(\dot{\tilde{v}}_i+F(\tilde{r})_i)\approx\nonumber\\\approx
-V_NF_0^{env}-({\nabla}F^{env}_{i0})\sum\limits_{i=1}^{N}\tilde{v}_i\tilde{r}_i\label{eqn2}
\end{eqnarray}

In the right-hand side of equation (2) the force $F_0^{env}$ in the
first term depends on $R$. It is a potential force. The second term
depending on coordinates of material points and their velocities relative
to the CM of the system determines changes in the internal energy
of the system. It is proportional to the divergence of the external
force. Therefore, in spite of the condition $R>>\tilde{r}_i$ the
values of $\tilde{v}_i$ may be not small, and the second term cannot
be omitted. Forces corresponding to this term are not potential
forces. So, the change in the internal
energy will be not equal to zero only if the characteristic scale of
inhomogeneities of the external field is commensurable with the
system scale.

Equation (2) confirms assumption of A. Poincare that it is
necessary to take into account structures of interacting bodies at
rather small distances between them [21].

Dynamics of an individual material point as well as dynamics of a system of
material points can be derived from equation (1). A material point
does not have an internal energy, and forces acting on it are caused
by potential forces of interaction with other material points and
the external force. Therefore the motion of a material point is
determined by the work of potential forces transforming the energy
of the external field and other material points into its kinetic
energy.

Unlike material points, a system has its internal energy. Therefore
the work of external forces over the system causes changes in its
$T_N^{tr}$ and $E_N^{ins}$, i.e. the external force breaks up into
two components. The first component is a potential force. It changes
momentum of the CM. The second component is non-potential. Its work
changes $E_N^{ins}$. Hence, the motion of the system is determined
by the work of potential and non-potential forces transforming the
external field energy into the energy of CM motion and internal
energy.

Multiplying eq.(1) by $V_N$ and dividing by $V_N^2$ we find the
equation of system motion [27]:
\begin{eqnarray}
M_N\dot{V}_N= -F^{env}-{\alpha_N}V_N\label{eqn3}
\end{eqnarray}

where $\alpha_N=[{\dot E}_N^{ins}+\Phi^{env}]/V_N^2$  is a
coefficient determined by the change of internal energy.

The equation (3) is a motion equation for ES. The first term in the
right-hand side of the equation determines the system acceleration,
and the second term determines the change of its internal energy.
The motion equation for ES is reduced to the Newton equation if it
is possible to neglect variation in the internal energy.

Thus, the system state in the external field is determined by two
parameters: the energy of motion and the internal energy. Each type
of energy has its own force. The change in the motion energy is
caused by the potential component of the force, whereas the change
in the internal energy is caused by the non-potential component.

It is known from the kinetics that the nonequilibrium system in
approach of the local equilibrium can be presented as a set of ES in
motion relative to each other [15]. In this case dynamic processes
in nonequilibrium systems will be defined by local values of energy
of ES relative motion and their internal energy. Therefore in this
case the description of dynamics of system by means of the equation
of ES motion (3) can be carried out if the external forces in it
equation to replace with the forces between ES.

Let us show how to obtain the equation for interaction between two
ES. For this purpose we take the system
consisting of two ES-$L$ and $K$.
 $L$ is the number of elements in the $L$-ES and $K$ is the number of elements in $K$-ES,
i.e. $L+K=N$. Let $LV_L+KV_K=0$, where $V_L$ and $V_K$ are
velocities of $L$ and $K$ equilibrium subsystems relative to the CM
system. Differentiating the energy of the system with respect to
time, we obtain:
${\sum\limits_{i=1}^{N}v_i{\dot{v}}_i}+{\sum\limits_{i=1}^{N-1}}\sum\limits_{j=i+1}^{N}v_{ij}
F_{ij}=0$, where $F_{ij}=U_{ij}=\partial{U}/\partial{r_{ij}}$.

In order to derive the equation for $L$-ES, in the left-hand side of
the equation we leave only terms determining change of kinetic and
potential energy of interaction of $L$-ES elements among themselves.
All other terms we displace into the right-hand side of the equation
and combine the groups of terms in such a way that each group
contains the terms with identical velocities. In accordance with
Newton equation, the groups which contain terms with velocities of
the elements from $K$-ES are equal to zero. As a result the
right-hand side of the equation will contain only the terms which
determine the interaction of the elements $L$-ES with the elements
$K$-ES. Thus we will have: ${\sum\limits_{i_L=1}^{L}}v_{i_L}
{\dot{v}}_{i_L}+{\sum\limits_{i_L=1}^{L-1}}\sum\limits_{j_L=i_L+1}^{L}
F_{{i_L}{j_L}}v_{{i_L}{j_L}}={\sum\limits_{i_L=1}^{L}}\sum\limits_{j_K=1}^{K}
F_{{i_L}{j_K}}v_{j_K}$ where double indexes are introduced to denote
that a particle belongs to the corresponding system. If we make
substitution $v_{i_L}=\tilde{v}_{i_L}+V_L$, where $\tilde{v}_{i_L}$
is the velocity of $i_L$ particle relative to the CM of $L$ -ES, we
obtain the equation for $L$-ES. The equation for $K$-ES can be
obtained in the same way. The equations for two interacting systems
can be written as [26]:
\begin{eqnarray}
V_LM_L\dot{V}_L+{\dot{E}_L}^{ins}=-{\Phi}_L-V_L{\Psi}
\end{eqnarray}
\begin{eqnarray}
V_KM_K\dot{V}_K+{\dot{E}_K}^{ins}={\Phi}_K+V_K{\Psi}
\end{eqnarray}

Here $M_L=mL, M_K=mK, \Psi=\sum\limits_{{i_L}=1}^LF^K_{i_L}$;
${\Phi}_L=\sum\limits_{{i_L}=1}^L\tilde{v}_{i_L}F^K_{i_L}$;
${\Phi}_K=\sum\limits_{{i_K}=1}^K\tilde{v}_{i_K}F^L_{i_K}$;
$F^K_{i_L}=\sum\limits_{{j_K}=1}^KF_{i_Lj_K}$;
$F^L_{j_K}=\sum\limits_{{i_L}=1}^LF_{i_Lj_K}$;
${\dot{E}_L}^{ins}={\sum\limits_{i_L=1}^{L-1}}\sum\limits_{j_L=i_L+1}^{L}v_{i_Lj_L}
[\frac{{m\dot{v}}_{i_Lj_L}}{L}+\nonumber\\+F_{i_Lj_L}]$;
${\dot{E}_K}^{ins}={\sum\limits_{i_K=1}^{K-1}}\sum\limits_{j_K=i_K+1}^{K}v_{i_Kj_K}
[\frac{{m\dot{v}}_{i_Kj_K}}{K}+\nonumber\\+F_{i_Kj_K}]$.

The equations (4, 5) are equations for interactions between ES.
They describe energy exchange between ES. Independent
variables are macro-parameters and micro-parameters.
Macro-parameters are coordinates and velocities of the motion of
CM of systems. Micro-parameters are relative coordinates
and velocities of material points.

Therefore the equation of system interaction binds together two
types of description: on the macrolevel and on the microlevel. The
description on the macrolevel determines dynamics of an ES as a
whole and description on the microlevel determines
dynamics of the elements of an ES.

The potential force, $\Psi$, determines the motion of an ES as a whole.
This force is the sum of potential forces acting
on the elements of one ES from the other system.

The forces determined by terms ${\Phi}_L$ and ${\Phi}_K$  transform
the motion energy of ES into their internal energy
as a result of chaotic motion of elements of one ES
in the field of forces of the other ES. As in the
case of the system in the external field, these terms are not zero
only if the characteristic scale of inhomogeneity of forces of one
system is commeasurable with the scale of the other system. The work
of such forces causes violation of time symmetry for ES dynamics.

The equations for systems motion corresponding to the equations
(4,5) can be written as:
\begin{equation}
M_L\dot{V}_L=-\Psi-{\alpha}_LV_L \label{eqn6}
\end{equation}
\begin{equation}
M_K\dot{V}_K=\Psi+{\alpha}_KV_K\label{eqn7}
\end{equation}
where ${\alpha}_{L}=(\dot{E}^{ins}_{L}+{\Phi}_{L})/V^2_{L}$,
${\alpha}_{K}=({\Phi}_{K}-\dot{E}^{ins}_{K})/V^2_{K}$,

The equations (6,7) are motion equations for two interacting ES.
The second terms in the right-hand side of the equations
determine the forces changing the internal energy of the ES.
These forces are equivalent to the friction forces. Their
work is a sum of works of forces acting on the material points of
one ES from the other ES.

The coefficients "$\alpha_L$", "$\alpha_K$" determine efficiency of
transformation of the energy of ES motion into their internal
energy. These coefficients are friction coefficients. Therefore
equations (6, 7) enable to determine analytical form of
non-potential forces in the non-equilibrium system causing changes
in the internal energy of the equilibrium system.

\section{The generals of Lagrange, Hamilton and Liouville equations for a set of equilibrium systems}

Let us show qualitative difference of Lagrange, Hamilton and
Liouville equations for the systems of material points from similar
equations for a system which consists from a set of interacting ES.

Using Newton equation one can derive Hamilton principle for material
points from differential D'Alambert principle [28, 29]. For this
purpose the time integral of virtual work done by effective forces
is equated to zero. Integration over time is carried out provided
that external forces possess a power function. It means that the
canonical principle of Hamilton is valid only for cases when $\sum
F_i\delta R_i=-\delta U$, where $i$ is a particle number, and $F_i$
is a force acting on this particle. But for interacting ES the
condition of conservation of forces is not fulfilled because of the
presence of a non-potential component. Therefore Hamiltonian
principle for a set of ES as well as Lagrange, Hamilton and
Liouville equations must be derived using motion equation for ES.

The equation for distribution function for a set of ES is
written as [25]:
\begin{equation}
df/dt=-\sum\limits_{L=1}^{R}{\partial}{F_L}/{\partial}V_L
\label{eqn8}
\end{equation}

Here $f$ is a distribution function for a set of ES,
$F_L$ is a non-potential part of collective forces
acting on the ES, $V_L$ is the velocity of
$L$-ES.

We call the equation (8) as Liouville equation for distribution
function of ES. Like the well-known Liouville equation for potential
interaction  of material points, the equation (8) is derived from
D'Alambert equation with the equation of motion for equilibrium
subsystems substituted into D'Alambert equation instead of Newton
equation for material points [28]. So, Liouville equation for ES
distribution function gets a nonzero right-hand side due to changes
in internal energy of equilibrium subsystems. Thus, equation (8)
actually defines the distribution function for particles with
internal degrees of freedom.

The right-hand side of the equation is determined by the efficiency
of transformation of the ES motion energy into their internal
energy. For non-equilibrium systems the right-hand side is not equal
to zero because of non-potentiality of forces changing the internal
energy.

The state of the system as a set of ES can be
defined in the phase space which consists of $6R-1$ coordinates and
momentums of ES, where $R$ is the number of
ES. Location of each ES is
given by three coordinates and their moments. Let us call
this space an $S$-space for ES in order to
distinguish it from the usual phase space for material points.
Unlike the usual phase space the $S$-space is not conserved. It is
caused by transformation of the energy of relative motion in
ES into their internal energy. The internal
energy cannot be transformed into the energy of motion as ES
momentum cannot change due to the motion of its material points [8].
Therefore $S$-space is compressible.

\section{The equations of interaction of systems and thermodynamics}

Equations (1-8) give relationship between mechanics and
thermodynamics [15,26]. According to the basic equation of
thermodynamics the work of external forces acting on the system
splits into two parts. The first part corresponds to reversible
work. In our case it corresponds to the change of the motion energy
of the system as a whole. The second part of energy goes on heating.
It corresponds to the internal energy of the system.

Let us take a motionless non-equilibrium system consisting of "$R$"
ES. Each ES consists of a great number of elements $N_L>>1$, where
$L=1,2,3...R, N=\sum\limits_{L=1}^{R}N_L$. In this case it is
possible to define the temperature, $T$ for ES. It is average
kinetic energy of a material points of ES. Let $dE$ be work done
over the system. In thermodynamics energy $E$ is called internal
energy (in our case it is equal to the sum of all energies of ES).
It is known from thermodynamics that ${dE=dQ-PdY}$. Here, according
to generally accepted terminology, $E$ is the energy of the system;
$Q$ is the thermal energy; $P$ is the pressure; $Y$ is the volume.
The equation of interaction between two systems is also a
differential of two types of energy. It means that $dE$ in the ES is
redistributed in such a way that some part of it changes energy of
relative motion of the ES and the other part changes the internal
energy. Thus, it follows that entropy may be introduced into
classical mechanics if it is considered as a quantity characterizing
increase in the internal energy  of an ES at the expense of energy
of their motion. Then the increase in entropy can be written as [26,
27]:
\begin{equation}
{{\Delta{S}}={\sum\limits_{L=1}^R{\{{N_L}
\sum\limits_{k=1}^{N_L}\int[{\sum\limits_s{{F^{L}_{ks}}v_k}/{E^{L}}]{dt}}\}}}}\label{eqn8}
\end{equation}

Here ${E^{L}}$ is the kinetic energy of $L$-ES; $N_L$ is the number
of elements in $L$-ES; $L=1,2,3...R$; ${R}$ is the number of ES;
${s}$ is the number of external elements which interact with ${k}$
element belonging to the $L$-ES; ${F_{ks}^{L}}$ is the force acting
on the $k$-element; $v_k$ is the velocity of the $k$- element.

Based on the generally accepted definition of entropy we can derive
expression for its production and define necessary conditions for
stationarity of a non-equilibrium system [27].

\section{Conclusion}
The key idea which enables for us to submit the explanation of
irreversibility without usage of the hypothesis of random
fluctuations is the idea that all surrounding objects have
structure. In other words all bodies consist of the structured
particles. The energy of the structured particle consists of two
principally different types. It is the energy of particle motion in
the external field and its internal energy. When particles interact
with each other, both types of their energies change. However, these
changes are different. Thus, kinetic energy of motion of a particle
changes as a result of work of potential component of the external
force. The internal energy changes as a result of work of
non-potential component of the force of interacting particles
transforming the energy of their motion into internal energy.
Therefore irreversibility is related to the structureness of
interacting bodies.

We describe transformation of both types of energies in the
framework of the model of the system consisting of ES of potentially
interacting material points. Such model enables to connect
micro-processes and macroprocesses of energy transfer during
interaction of ES. Indeed, the equation for motion of ES includes
microparameters-coordinates and velocities of material points of
which ES are composed, whereas macroparameters include coordinates
and velocities of ES.

The evolution of closed non-equilibrium system represented by a set
of ES is determined by potential and non-potential forces acting
between ES. Potential forces change kinetic energy of motion of ES.
The work of non-potential forces transforms energy of motion of an
ES into its internal energy. The phase space determined by
coordinates and velocities of such system we called $S$-space. It is
compressible. Compression of $S$-space is determined by Liouville
equation for non-equilibrium system from a set of ES. The system
acquires an equilibrium state when all energy of ES motion
transforms into its internal energy.

The obtained equations for structured particles give relationship
between classical mechanics and thermodynamics. Therefore according
to the motion equation for ES the first law of thermodynamics follows
from the fact that the work of external forces changes both the
energy of particle motion and their internal energy. The second law
of thermodynamics follows from irreversible transformation of energy
of relative motion of system's particles into their internal energy.

The motion equation for ES also states impossibility of existence of
structureless particles in classical mechanics, which is equivalent
to infinite divisibility of matter.

As an example of importance of the equation of motion for structured
particles we should note the following. In strong interactions the
internal degrees of freedom of interacting particles are exited.
According to the equation of motion for structured particles it
causes changes in their internal energy and corresponding violation
of time symmetry. It is impossible to take into account these facts
only using canonic Hamiltonian formalism without any empirical
corrections.

The offered explanation of irreversibility for the structured
particles is applicable for nonequilibrium systems only at
possibility of their representation by model in the form of set of
ES. The clusters and other structures presence in continuous
environments tell us about generality of such model [30].
Nevertheless, it is necessary to investigate in future the question
about how this condition limits applicability of the offered
mechanism of irreversibility.

\smallskip

\end{document}